%% file: Template.tex
\documentclass{article}
\usepackage{spconf,amsmath,graphicx}
\usepackage{multirow}
\usepackage{booktabs}
\usepackage{url}
\usepackage{array}
\usepackage{longtable}
\usepackage{amsmath}
\usepackage{mathtools,amssymb}
\input{math_commands.tex}

\newcolumntype{C}[1]{>{\centering\arraybackslash}m{#1}}

\title{DurIAN-E 2: Duration Informed Attention Network with Adaptive Variational Autoencoder and Adversarial Learning for Expressive Text-to-Speech Synthesis }
\name{Yu Gu, Qiushi Zhu, Guangzhi Lei, Chao Weng, Dan Su}
\address{Tencent AI Lab}
\begin{document}
%
\maketitle
\begin{abstract}
This paper proposes an improved version of DurIAN-E (DurIAN-E 2), which is also a duration informed attention neural network for expressive and high-fidelity  text-to-speech (TTS) synthesis.  Similar with the DurIAN-E model,
multiple stacked SwishRNN-based Transformer blocks are utilized as linguistic encoders and Style-Adaptive Instance Normalization (SAIN) layers are also exploited into frame-level encoders to improve the modeling ability of expressiveness in the proposed the DurIAN-E 2.
Meanwhile, motivated by other TTS models using generative models such as VITS,  the proposed DurIAN-E 2 utilizes variational autoencoders (VAEs) augmented with normalizing flows and a BigVGAN waveform generator with adversarial training strategy, which further improve the synthesized speech quality and expressiveness.  Both objective test and subjective evaluation results prove that the proposed expressive TTS model DurIAN-E 2 can achieve better performance than several state-of-the-art approaches besides  DurIAN-E.
\end{abstract}
\section{Introduction}
\label{sec:intro}
In recent years,  many state-of-the-art TTS systems  based on deep neural networks have made rapid progress and were able to synthesize more natural and high-quality speech,
compared with conventional unit selection  concatenative and statistical parametric speech synthesis approaches. Although those TTS systems have synthesized speech qualitatively similar to real human speech, there still exists a huge gap between TTS-synthetic speech and human speech in terms of expressiveness.
Expressive TTS technology which  aims to model and control the speaking style  has become a hot research topic  for decades.
 At present, there are 
two mainstream approaches to model the speaking style information:
one uses pre-defined categorical style labels as the global control condition of TTS systems to denote different speaking styles \cite{tits2019visualization} and the other imitates the speaking style given a reference speech \cite{wang2018style, skerry2018towards}. For the first kind of approach, the style control strategy is more intuitive and interpretable, which is more suitable for practical TTS applications. For the second one, the global style tokens or style embeddings extracted from the training datasets can enrich the diversity of expressiveness and additional style labels are not required. 

DurIAN-E \cite{duriane} was an auto-regressive (AR) expressive TTS model, in which the alignments between the input linguistic information and the output acoustic features were inferred from a duration model.
Inherited from DurIAN \cite{yu2020durian}, DurIAN-E could generate mel-spectrograms  from linguistic features frame by frame using AR decoders \cite{tacotron2} given pre-defined rich categorical style labels.  To improve expressiveness, 
SAIN layers \cite{li2022styletts} were also employed to learn the distribution with a style-specific mean and variance for each channel in mel-spectrograms  and  each channel in the mel-spectrograms represented a single frequency range.
SwishRNN \cite{lei2022simple},  an extremely simple recurrent module, was built to substitute the feed-forward blocks in Transformer \cite{transformer}, which could improve pronunciation stability. Denoising diffusion  probabilistic models (DDPMs) \cite{ddpm} with SAIN layers were also applied as the denoisers, which enhanced the output mel-spectrograms generated from the AR decoders and could achieve better speech quality and expressiveness. 

However DurIAN-E was also a two-stage model, which predicted acoustic features from the linguistic information  using an acoustic model and generated speech waveforms through an additional neural vocoder separately. These two-stage models usually suffer from the critical mismatch of different distributions between acoustic features predicted from the acoustic model and those used to train the neural vocoders,  
which can lead to artifacts in the synthesized speech. To address this issue, many TTS models combine acoustic models and vocoders by learning a latent representation with certain meanings and generate waveforms directly from linguistic information \cite{gu2018multi, glowwavgan, vits}. Motivated by these fully end-to-end models such as Glow-TTS \cite{glowtts} and VITS \cite{vits}, we improve DurIAN-E by replacing the AR decoder and the DDPM-based denoiser with a BigVGAN decoder \cite{bigvgan} to tackle the mismatch problem and increase the inference efficiency.
Normalizing flows conditioned on style labels and  hidden sequences from prior encoder
are employed to model the latent representation learned from linear spectrograms by SAIN-based VAEs. 

 \begin{figure*}[t]
  \centering
  \includegraphics[width=\linewidth]{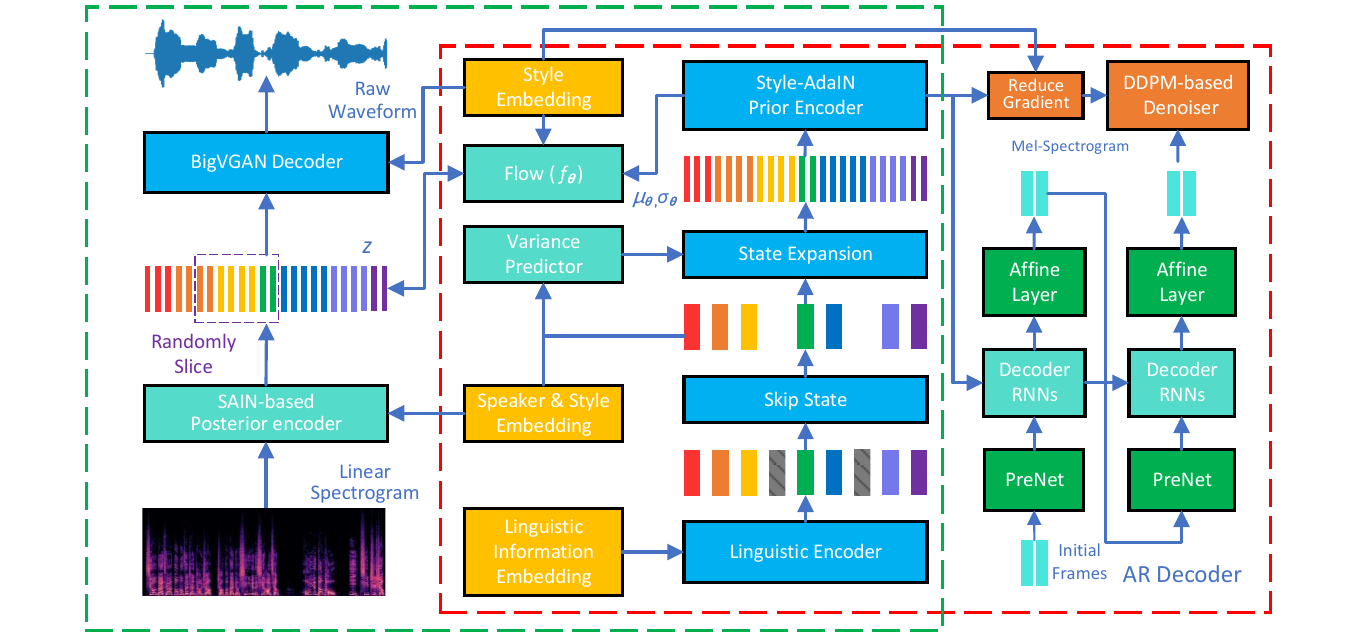}
  \caption{Model structures of DurIAN-E (shown in red dashed box) and DurIAN-E 2 (shown in green dashed box).}
  \label{fig:structure}
\end{figure*}
 
 This paper is organized as follows. 
Section \ref{sec:duriane2} introduces the proposed  DurIAN-E 2 model in this paper and the constructed expressive TTS system in detail. The experimental conditions and results are described in  Section \ref{sec:exp} and finally Section \ref{sec:con}  concludes this paper.

\section{DurIAN-E 2}
\label{sec:duriane2}
\subsection{Architecture}
The model structure of the proposed DurIAN-E 2 is illustrated in the green dashed box of Fig.  \ref{fig:structure}. In order to better reflect the relationship and differences between  the architectures of DurIAN-E and DurIAN-E 2,  the model structure of the original DurIAN-E is also displayed in the red dashed part of Fig. \ref{fig:structure}. DurIAN-E 2 
can be regarded as a VAE whose prior encoder is conditioned on
linguistic information and posterior encoder is conditioned on 
acoustic features.
The overall architecture of DurIAN-E 2  consists of  variance predictors, a state-skip prior encoder, SAIN-based posterior encoder, BigVGAN  decoder and discriminators. 
DurIAN-E outperformed several state-of-the-art TTS systems on subjective tests \cite{duriane}, therefore
 some proposed structures of DurIAN-E are retained completely in DurIAN-E 2. 
Similar with DurIAN-E, DurIAN-E 2 also possesses the same variance predictors as DurIAN-E,  which can predict  duration, pitch and pitch range of each phoneme. To accelerate the convergence and promote stability of model training,  the ground truth duration of each phoneme is obtained through forced alignment using GMM-HMMs rather than using monotonic alignment search like VITS and Glow-TTS.
Besides the modules inherited from DurIAN-E directly, the others in DurIAN-E 2 are inspired by other TTS works \cite{vits, bigvgan}, in which
the posterior encoder and discriminator are only adopted for training. The architectural details are introduced as follows. 

\subsection{Prior encoder}

\begin{figure}[t]
  \centering
  \includegraphics[width=\linewidth]{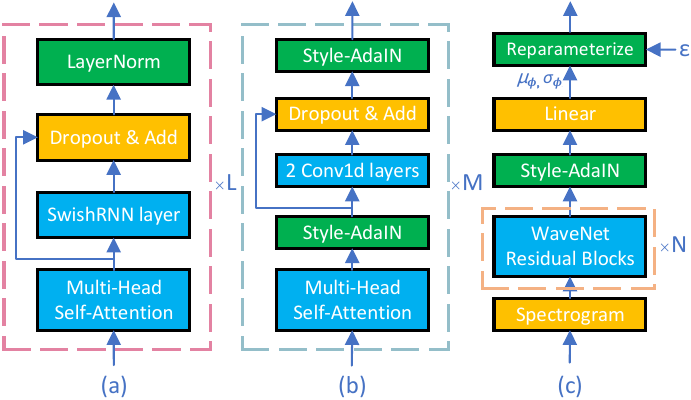}
  \caption{Blocks used in the encoders of DurIAN-E 2, 
  where $\mu$, $\sigma$ and $\epsilon$ represent mean, variance and Gaussion noise vectors. }
  \label{fig:block}
  \vspace{-0.1cm}
\end{figure}

 The two-level hierarchical skip-encoders exploited in DurIAN-E can significantly reduce pronunciation errors and improve speech quality and expressiveness, which consist of a phoneme-level linguistic encoder with a stack of SwishRNN-based Transform blocks \cite{lei2022simple} as displayed in Fig.~\ref{fig:block} (a) and a frame-level encoder with multiple convolutional Transformer blocks and SAIN-based layers as displayed in  Fig.~\ref{fig:block} (b). 
 This two-level hierarchical skip-encoder  structure is also reserved to DurIAN-E 2.
Different with DurIAN-E, we treat the skip-encoders as the prior encoders of VAE and  a linear projection layer above the skip encoder is conducted to produce the mean and variance  values used for constructing the prior distribution. 
 A normalizing flow $f_{\theta}$ with a stack of affine coupling layers \cite{dinh2016density} and WaveNet residual blocks \cite{wavenet} is attached to improve the flexibility of the prior distribution. The factorized normal prior distribution is complexified with a set of invertible transformations as following:
\begin{align}
\label{equ:prior}
    p_\theta(\vz|\vc) & = N(f_{\theta}(\vz);\mu_{\theta}(\vc),\sigma_{\theta}(\vc))\Big|\det\frac{\partial f_{\theta}(\vz)}{\partial \vz}\Big|, 
\end{align}
where $\vc$ is the output sequence of the frame-level SAIN-based encoder.
Style embedding  is added to each WaveNet residual block in the normalizing flow as the global condition. 

\subsection{SAIN-based posterior encoder}
For the posterior encoder which approximates the posterior distribution $q_\phi(\vz|\vx)$ of the latent variables $\vz$ given acoustic feature condition $\vx$, the same non-causal WaveNet residual block consisting of layers of dilated convolutions with a gated activation unit and skip connection as Glow-TTS is used as shown in Fig.~\ref{fig:block} (c).  Style embeddings from the categorical style labels are added to the residual blocks as the global condition \cite{wavenet}.
To better distinguish different acoustic features from multiple styles and further improve effectiveness, a SAIN layer which is also deployed in prior encoder and an additional linear projection layer are also added above the WaveNet blocks to produce the mean and variance of the normal posterior distribution.
The KL divergence is then calculated to optimize the model:
\begin{align}
\label{eqn/loss_kld}
   \mathcal{L}_{KL} = \log{q_{\phi}(\vz|\vx)} - \log{p_{\theta}(\vz|\vc)},
\end{align}
where   $ \vz \sim q_{\phi}(\vz|\vx) = N(\vz;\mu_{\phi}(\vx),\sigma_{\phi}(\vx))$ and $\vx$ is the high-resolution  linear-scale spectrogram of target speech to provide more acoustic details.

\subsection{BigVGAN decoder}
\label{ssec:bigvgan}
Different with VITS, we employed BigVGAN generator \cite{bigvgan} as the waveform decoder conditioned on the latent hidden variables $\vz$ in the VAE  rather than HiFi-GAN \cite{hifigan}. 
The generator is composed of multiple blocks of transposed convolution followed by the anti-aliased multi-periodicity composition (AMP) module. 
The AMP module adds features from multiple residual blocks with different channel-wise periodicities before dilated 1-D convolutions, which uses Snake function for providing periodic inductive bias, and low-pass filter for anti-aliasing purpose. As the global condition, the style embedding is added to the input latent representation $\vz$.

We follow the discriminator architectures of BigVGAN, using the multi-resolution discriminators (MRDs) on the time–frequency domain, which operates on multiple 2-D linear spectrograms with different STFT resolutions and using the multi-period discriminator (MPDs),  where the 1-D signal is reshaped to 2-D representations with varied heights and widths to separately capture the multiple periodic structures though 2-D convolutions.
Following objectives $\mathcal{L}_G$ for generator and $\mathcal{L}_D$ for discriminator are calculated respectively:
\begin{align}
 \mathcal{L}_G  =\sum_{k=1}^{K}\bigg[\mathcal{L}_{adv}(G;D_k)+\lambda_{fm}\mathcal{L}_{fm}(G;D_k)\bigg] \notag \\
 + \lambda_{mel}\mathcal{L}_{mel}(G), \quad \mathcal{L}_D  = \sum_{k=1}^{K}\bigg[\mathcal{L}_{adv}(D_k;G)\bigg],   
\end{align}
where $D_k$ denotes the $k$-th MPD or MRD discriminator submodules. $\mathcal{L}_{adv}$ uses the least-square GAN loss, $\mathcal{L}_{fm}$ is the feature matching loss \cite{kumar2019melgan} which minimizes the $\ell_1$ distance for every intermediate features from the discriminator layers and 
$\mathcal{L}_{mel}(G)$ denotes the spectral $\ell_1$ regression loss between the mel-spectrogram of the synthesized waveform and the corresponding ground-truth.

\begin{table}[t]
 \centering
\renewcommand{\arraystretch}{1.1}
  \caption{Distortion between acoustic features of natural and synthesized speech from different systems.
    V/UV  means frame-level voiced/unvoiced error. BAP and Corr. denote the
BAP prediction error and correlation coefficients respectively.}
  
    \setlength{\tabcolsep}{1.5mm}{
      \begin{tabular}{ c |ccccc}
       \toprule
        \multirow{2}{*}{System} & MCD &  BAP & F0 RMSE  & F0 & V/UV  \\
          &  (dB) & (dB) & (Hz)& Corr. &(\%)\\
        \midrule
        \emph{\textbf{GT (vocoder)}}  & 2.898 & 2.708 & 15.765 &0.975&3.863 \\
       \midrule
       \emph{\textbf{FastSpeech2}}& 6.138& 4.274 & 48.921   & 0.758&7.603\\
        \emph{\textbf{DurIAN}}  & \textbf{6.036} & 3.851 & 47.908& 0.769& \textbf{7.334}\\
       \midrule
          \emph{\textbf{DiffSpeech}}& 6.711& 3.361 & 50.249  & 0.749&8.205\\
        \emph{\textbf{DurIAN-E}} & 6.686 & 3.354 & 48.469 & 0.765 & 7.440\\
       \midrule
       \emph{\textbf{VITS}} & 6.691 & 3.285 & 49.420 & 
       0.751 & 8.235 \\
      \emph{\textbf{DurIAN-E 2}} & 6.500 & \textbf{3.281} & \textbf{47.899} & \textbf{0.773} & 7.550\\
      \bottomrule
      \end{tabular} }    

       \label{tab:obj}
  \end{table}

\begin{table*}[t]

    \small

    \begin{tabular}{c| cc| cccc|c c}
       \toprule
   System & \emph{\textbf{GT}} & \emph{\textbf{GT (vocoder)}} & \emph{\textbf{FastSpeech 2}} &  \emph{\textbf{DurIAN}} &  \emph{\textbf{DiffSpeech}}&  \emph{\textbf{VITS}} & \emph{\textbf{DurIAN-E}} & \emph{\textbf{DurIAN-E 2}}\\
   \midrule
       \textbf{MOS} & 4.45 $\pm$ 0.16 & 4.23 $\pm$ 0.17 & 3.62 $\pm$ 0.17 &  3.73 $\pm$ 0.15 & 3.78 $\pm$ 0.19  & 3.80 $\pm$ 0.23 & 3.86 $\pm$ 0.14  & \textbf{4.00 $\pm$ 0.18} \\
            \bottomrule
    \end{tabular}
    \centering
      \caption{The MOS values of different systems with 95\% confidence intervals.}
       \label{tab:mos}
\end{table*}

\section{Experiments}
\label{sec:exp}
To better evaluate the performance of the proposed DurIAN-E 2 model with DurIAN-E and other TTS models,\footnote{Examples of synthesized speech by different systems are available at \url{https://sounddemos.github.io/durian-e2}.} the identical experimental setup to DurIAN-E \cite{duriane} was adopted, in which a high-expressiveness Chinese corpus containing 11.8 hours of speech with 12 kinds of styles pronounced by 7 different speakers was used as the training dataset.
TTS systems including  \textit{\textbf{DurIAN}} \cite{yu2020durian},  \textit{\textbf{FastSpeech 2}} \cite{renfastspeech},  \textit{\textbf{DiffSpeech}} \cite{liu2022diffsinger}, \textit{\textbf{VITS}} \cite{vits} and \textit{\textbf{DurIAN-E}} \cite{duriane} were established for comparison. 
 We also completed the objective tests and mean opinion score (MOS) tests on the basis of previous experiments in \cite{duriane}. Note that
all these systems except VITS and the proposed DurIAN-E 2 shared an extra unified BigVGAN vocoder \cite{bigvgan} which was trained separately conditioned on ground truth mel-spectrograms to better compare the performances among different acoustic models.
\label{ssec:obj}
\subsection{Objective tests}
Objective tests were conducted to evaluate different synthesis systems. In this test, all  systems remained the same ground-truth duration as the target natural speech for the convenience of comparison.
Mel-cepstral distortion (MCD),  distortion of band aperiodicities (BAP), voiced/uvoiced prediction error,  root-mean-square error (RMSE) and correlation coefficients of F0 values on a linear scale  between natural speech and synthesized speech by different systems are presented in Table \ref{tab:obj} and these systems are divided into three groups according to whether they employed DDPM-based denoisers or generated waveforms directly. The compared acoustic features  were re-extracted from the synthesized waveforms. 

\begin{figure}[t]
  \centering
  \includegraphics[width=0.95\linewidth]{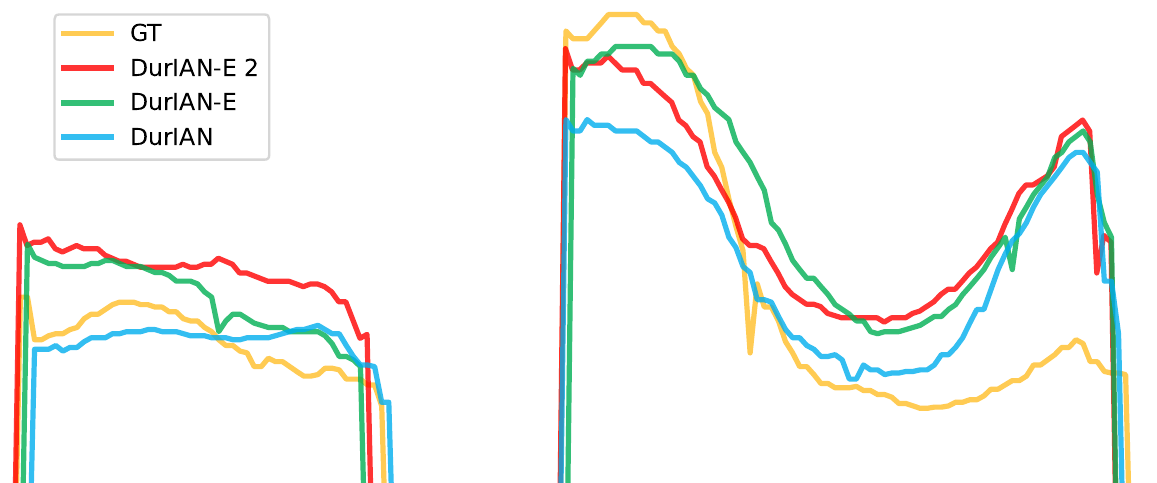}
  \caption{Illustration of pitch contours from different systems.}
  \label{fig:pitch}
\end{figure}

\begin{figure}[t]
  \centering
  \includegraphics[width=\linewidth]{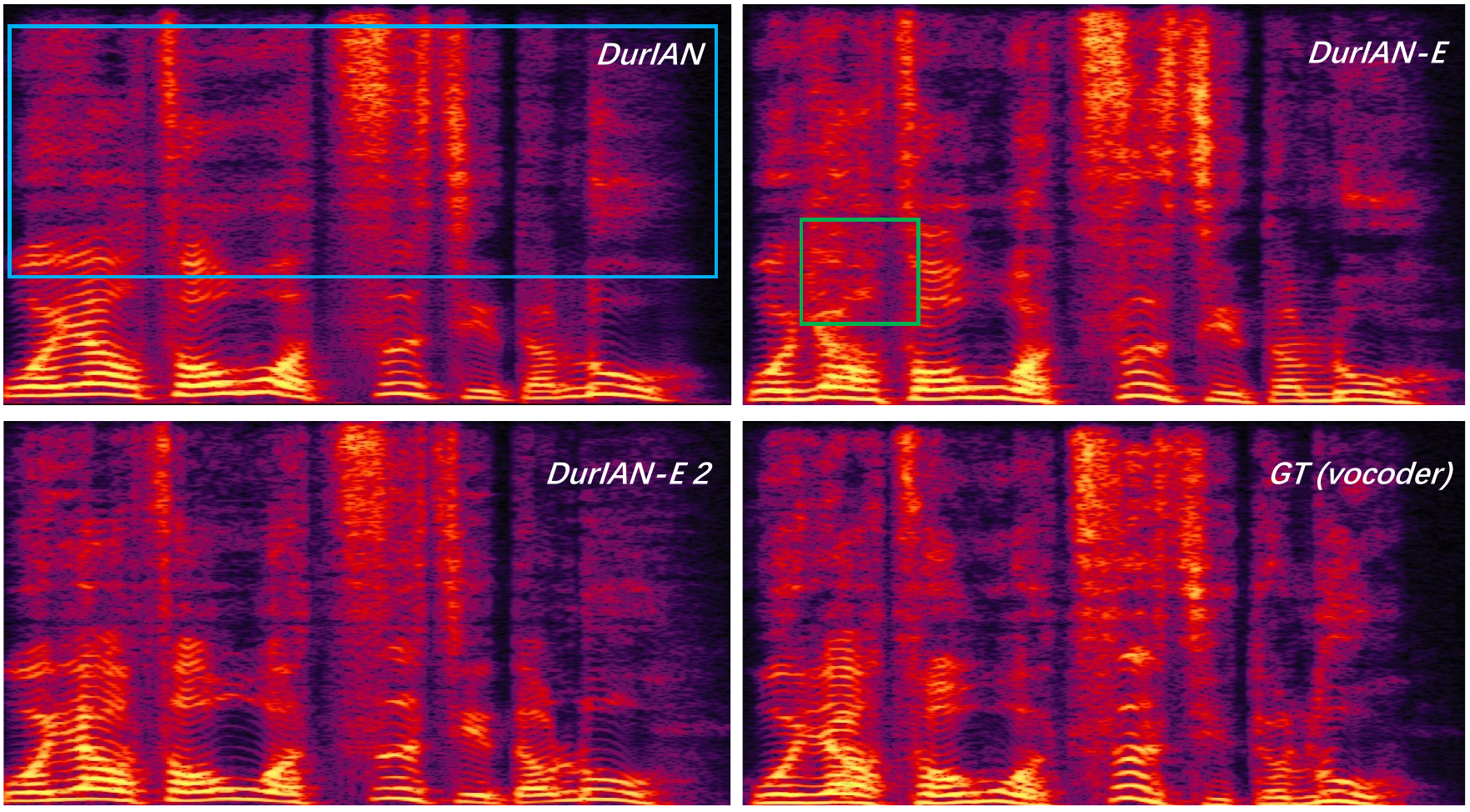}
  \caption{Illustration of spectrograms from different systems.}
  \vspace{-0.5cm}
  \label{fig:spg}
\end{figure}

The objective results show that the synthesized speech from \textit{\textbf{DurIAN-E 2}} system can obtain the most accurate F0 and BAP values.  Fig. \ref{fig:pitch} gives an example of pitch contours from different DurIAN-related systems.
The F0 curve of \emph{\textbf{DurIAN}}  is much smoother and lacks fluctuations than \emph{\textbf{DurIAN-E}} and \emph{\textbf{DurIAN-E 2}}. The MCD value in \emph{\textbf{DurIAN-E 2}} is also smallest among the systems using DDPM-based denoisers and  those generating waveforms directly. The spectral distortions in the group with 
\emph{\textbf{DurIAN}} and \emph{\textbf{FastSpeech 2}} are slightest among three groups mainly because the acoustic models in these systems directly optimized the MSE losses of spectral features.
Although
 \emph{\textbf{DurIAN}} and  \emph{\textbf{FastSpeech 2}} can achieve the smallest MCD, the over-smoothing problems as depicted in Fig. \ref{fig:spg}  severely decrease speech quality.
  \emph{\textbf{DurIAN-E 2}} outperforms \emph{\textbf{DurIAN-E}} and  could generate clearer harmonic structures and more similar spectral details with the ground truth  as illustrated in Fig. \ref{fig:spg}.

\subsection{Subjective MOS tests}
\label{ssec:sub}
For each test, 20 test utterances randomly selected from the test set were synthesized by different systems and evaluated  by 10 experienced listeners.
MOS results are listed in Table \ref{tab:mos}. \emph{\textbf{DurIAN-E 2}}  can achieve the best MOS score among all TTS systems, which demonstrates  model capacity of the proposed system is sufficient. 
The system \emph{\textbf{VITS}} which can  directly generate waveforms in a fully end-to-end manner transcends the  multi-stage model \emph{\textbf{DiffSpeech}} in which 
predicted mel-spectrograms were firstly enhanced by DDPM-based denoisers and then converted to waveforms by vocoders. Although \emph{\textbf{DurIAN-E}} is also such a multi-stage model, due to the employment of the two-level hierarchical skip-encoder architecture,
 \emph{\textbf{DurIAN-E}} obtained better MOS result than   \emph{\textbf{VITS}}. \emph{\textbf{DurIAN-E 2}} 
 combines the advantages of those models, which consists of the end-to-end waveform generation method  in \emph{\textit{VITS}} and other complicated structures such as SwishRNNs and SAIN layers as the prior encoder and the positerior encoder and therefore can synthesize more expressiveness and higher fidelity speech.

\section{Conclusion}
\label{sec:con}
In this paper, we propose DurIAN-E 2, an improved model  for expressive and high-fidelity TTS based on style adaptive VAEs, which utilizes the two-level hierarchical skip-encoder architecture in DurIAN-E as the prior encoder and the SAIN-based posterior encoder to achieve more natural prosody and better expressiveness. BigVGAN decoder conditioned on variable hidden vectors and style embeddings is also employed to further improve the speech quality.
Experimental results of both objective and subjective tests prove that DurIAN-E 2 can achieve better performance than the state-of-the-art approaches besides DurIAN-E.


\vfill\pagebreak

\bibliographystyle{IEEEbib}
\bibliography{am,refs}

\end{document}

%% file: math_commands.tex
\usepackage{amsmath,amsfonts,bm}

\def\eqref#1{equation~\ref{#1}}

\def\1{\bm{1}}

\def\vc{{\bm{c}}}

\def\vx{{\bm{x}}}

\def\vz{{\bm{z}}}

\DeclareMathAlphabet{\mathsfit}{\encodingdefault}{\sfdefault}{m}{sl}
\SetMathAlphabet{\mathsfit}{bold}{\encodingdefault}{\sfdefault}{bx}{n}

